# Glycine in the interstellar medium revisited:
## Resolving conformational complexity kinetically

J. Wurmel[1,†] and J. M. Simmie[‡]

† Department of Analytical, Biopharmaceutical and Medical Sciences,

Atlantic Technological University, ATU, Galway H91 T8NW, Ireland

‡ School of Biological and Chemical Sciences, University of Galway H91 TK33, Ireland

## Abstract

Numerous attempts to detect glycine, a simple amino acid and a key building block of proteins in all life on Earth, in the interstellar medium have so far been unsuccessful in spite of the fact that it is present in meteorites, comets and asteroids. Glycine is conformationally diverse, existing in eight different forms, and the aim is to show that interconversions, in the dark, are governed by quantum mechanical tunnelling at temperatures from 300 K down to 10 K and to compute the rate at which these changes occur resulting in recommendations to maximise a successful outcome.

# Introduction

The organic compound glycine, $H_2NCH_2C(O)OH$, is one of 20 proteinogenic amino acids and a vital building block of proteins in all life on Earth. It has been definitively identified in comets and asteroids, which strongly suggests an extraterrestrial origin for prebiotic molecules. The synthesis of glycine in the interstellar medium is an active research area with many proposed formation pathways, (Mates-Torres, 2026), (Pinto,2025), (Choe, 2023), (Rimola, 2022), (Ioppolo, 2021), (Sato, 2018), (Garrod, 2013) none of which distinguish between the different forms, or conformers[2] of the molecule.

In a seminal paper (Lattelais, 2011) provided a set of guidelines to aid in the search for glycine in the interstellar medium, which until that time had not been detected, although it had been known to exist in carbonaceous meteorites (Engel, 1982) and comet samples (Glavin, 2008).

---

[1] Corresponding author: john.simmie@universityofgalway.ie

[2] Different spatial arrangements of the same molecule achieved by rotation about a single bond

They reviewed the large number of attempts that had been made to detect neutral glycine and enunciated a minimum energy principle to show that two isomers[3] of glycine, namely N-methylcarbamic acid $CH_3NHC(O)OH$ and methyl carbamate $CH_3OC(O)NH_2$, both part of the $C_2H_5O_2N$ family, were more likely to be detected given their higher stability and larger dipole moment.

The first isomer of glycine to be found (Rivilla, 2023) was neither of these but $HOCH_2C(O)NH_2$, glycolamide, detected towards the G+0.693-0.027 molecular cloud. A higher-energy conformer of glycolamide as well as two conformers of glycine were below the limit of detection which led the authors to conclude that explanations based on thermodynamic equilibria were unfeasible and that chemical kinetics arguments would be required to explain these findings.

More recently samples taken from asteroids Ryugu (Grady,2025) and Bennu (Glavin, 2025) have put the matter beyond all doubt, with glycine detected at the ppm level in surface and sub-surface samples and those from Bennu containing high abundances of amino acids with glycine dominant.

A number of explanations have been advanced by (Rivilla, 2023) to explain the anomaly such as that molecules with carboxyl groups -C(O)OH are rare, therefore making the formation of glycine less likely. By analogies with the observed ratios for related molecules such as ethanol :ethanolamine, CH3CH2OH :$H_2NCH_2CH_2OH$ of $\approx 40$, one could expect a not too dissimilar ratio for acetic acid : glycine, $CH_3C(O)OH$:$H_2NCH_2C(O)OH$. Using measured abundances of acetic acid, this reasoning suggests that glycine is likely present at levels below current detection limits.

Conformational diversity is not helpful as it dilutes the abundance of a particular species and increases spectral complexity which is exacerbated in nitrogen $^{14}N$ compounds by its non-zero nuclear quadrupole moment, $I = 1$; glycine which exists in some neutral eight forms (Csaszar, 1992), (Orjan, 2020} is a good example of this. In addition, only two conformers of glycine (*saa* and *ass*) have been studied by microwave spectroscopy, (Lovas, 1995), (McGlone, 1999), (Ilyushin, 2005) thus providing the most reliable data for searches. A rigorous set of conditions that must be met before a detection event for glycine can be deemed successful has been outlined (Snyder, 2005).

Detecting more than one conformer can provide insight into formation routes; for example, (Molpeceres, 2021) show that the preponderance of *trans*-thioformic acid, HC(O)SH, formed by successive hydrogenation of carbonyl sulfide, O=C=S, on amorphous solid water is due to the first step in which the *cis*-intermediate is preferentially generated, which in turn means that the *trans*-thioformic acid is favoured in the second hydrogenation step.

Laboratory measurements by (Bazso, 2012) have shown that the lowest energy conformer of glycine, *saa*, in Ar, Kr, Xe and $N_2$ matrices at 12~K can be transformed into the higher-energy conformer *aaa* by near-infrared laser irradiation (hv); in the dark the *aaa* then reverts to *saa* with half-lives[4], $\tau$, of between 2.8 and 4.4 s in the noble-gases and much longer times in dinitrogen, Fig 1. Thus, the rotamerisation 'reaction' rate constant, $k$, in which the H—O--C=O dihedral flips from 180º to 0º can be derived since $k = \ln(2) / \tau$. The authors were able to show that quantum mechanical tunnelling by the hydrogen atom is responsible since a conventional reaction, an over-the-energy-barrier process, is highly unlikely given that a barrier of 2,147 $cm^{-1}$ (3,090~K or 25.7 kJ $mol^{-1}$ is too high to overcome at these low temperatures. They employed a simple one-dimensional model (Eckart, 1930) to predict quantum mechanical tunnelling half-lives of 658 and 383 s at 12 and 15 K, respectively — times which are considerably slower than those observed. Additional experiments employing isotopic substitution confirmed that the reaction does in fact proceed exclusively by quantum mechanical tunnelling.

[3] Molecules that share the same molecular formula but possess different structures.

[4] The time required for a number of molecules to decrease to half of their initial value.

The glycine cryogenic experiments of (Bazso, 2012) have been simulated using semi-classical transition state theory and ONIOM to include the matrix with good agreement with experiment for Ar, Kr and Xe matrices by (Mandelli, 2026} but their gas-phase values consistently predict much larger half-lives than experiment.

# Methodology

The geometries, frequencies and energies of all eight known neutral conformers of glycine[5] were determined using the density functional B2PLYP supplemented by a dispersion correction D3 and Becke-Johnson damping BJ (Grimme, 2010) with the chemistry code Gaussian-16 (Gaussian, 2016) The resultant relative zero-point corrected electronic energies are in excellent agreement with the wave function theory values of (Orjan, 2020) obtained at the CCSD(T)/aug-cc-pVTZ level of theory.

Calculation of the chemical kinetics proceeds from the reactant, transition state and product structure files to generate a minimum energy path from whence variational transition state theory rate constants are evaluated including small-curvature and quantised states tunnelling approximations over a range of temperatures from 300 to 10 K using the application Pilgrim (Ferro-Costas, 2023).

# Results

It is generally agreed that glycine formation in the ISM takes place on water–ice grains and that the molecule desorbs into the gas-phase at temperatures of ≈ 200 K (Mates-Torres, 2026), (Pinto,2025), (Choe, 2023), (Rimola, 2022), (Ioppolo, 2021), (Sato, 2018), (Garrod, 2013); however, none of the postulated routes are conformer-specific, although Fig. 2 in (Rimola, 2026} appears to indicate that *ass* formation is favoured.

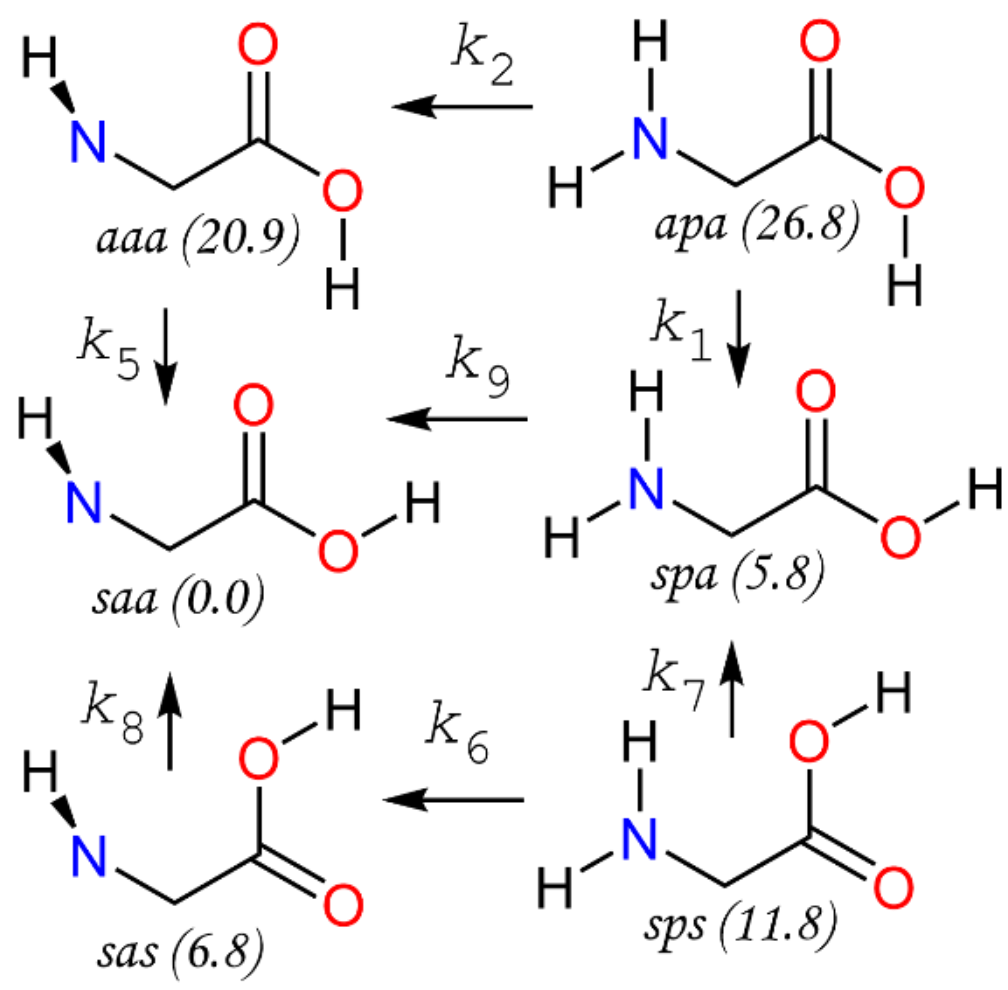

*Figure 1:Glycine conformers with relative energies (kJ mol-1): experiments saa to aaa by hv. Fate of conformers: apa either ⟶ spa ⟶ saa or ⟶ aaa ⟶ saa*

[5] The nomenclature is adopted from (Ruscic, 2025)

We assume that all eight glycine conformers are initially present initially in the ISM, and analyse the subsequent exergonic transformations that occur in the dark. In the previously studied matrix experiments the *aaa* conformer reverts to *saa* through a flip of the H-O-C=O dihedral from 180º, Fig.1. An analogous process occurs with the highest-energy conformer, *apa*, which connects to *spa*; subsequent rotation of the $H_2N$ group then yields to the lowest energy (ground-state) conformer, *saa*. Note that *apa* is identical to *aaa* except that the $H_2N$ group has rotated to face the viewer, and similarly *spa* is identical to *aaa* except for the rotated $H_2N$, Fig 1.

We can categorise the various transformations as (1) H-O-C=O dihedral flips or rotamerisations, (2) hydrogen-atom transfers from one oxygen atom to the other and (3) change in the orientation of $–NH_2$. Rotations about the C–C bond, which do interconvert conformers, can be discounted due to very high barriers.

These transformations account for five of the seven higher-energy conformers; the remaining two are outlined in Figure 2., which shows that *aas* reverts via *sas* to *saa* but reverting via *ass* results in a dead-end since this lowest-lying conformer is just 5.8 kJ mol$^{-1}$ above the ground-state and there are no possible routes by which it can become *saa* since it would involve three simultaneous changes (1) a H-atom transfer, (2) a H-O-C=O dihedral flip and (3) a $NH_2$ group flip.

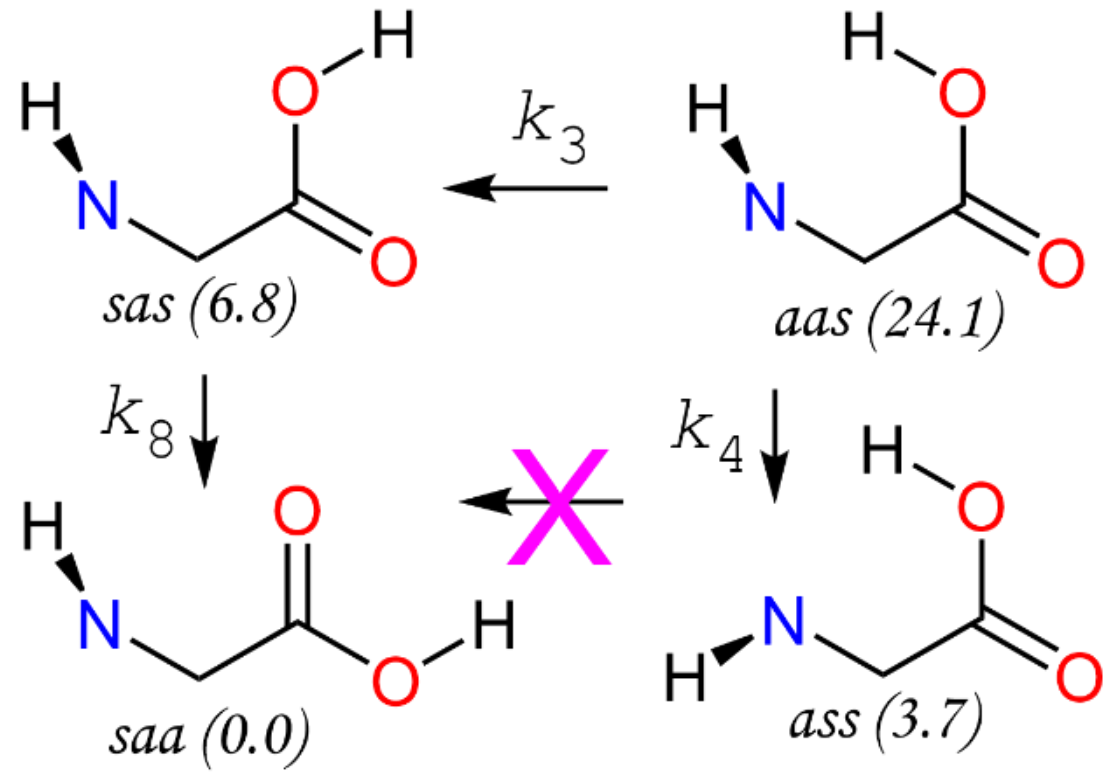

*Figure 2:* Fate of conformers *aas* ⟶ *sas* ⟶ *saa* or ⟶ *ass*

At 10 K, the temperature at which the matrix experiments were conducted, our computed value for the half-life of 6.4 ms for *aaa* ⟶ *saa* is considerably faster than the observed of 4 s — but it is known that the molecule–matrix interactions generally tend to slow the reaction increasing the half-life (Drabkin, 2025).

*Table 1: Rate constants, ki, and half-lives, ?, at 200 K*

| Reactions | $k$ / s$^{-1}$ | τ |
|---|---|---|
| *apa* →*spa* $k_1$ | 1.7E+06 | 0.41 μs |
| *aas* →*sas* $k_3$ | 5.6E+04 | 12 μs |
| *aaa* →*saa* $k_5$ | 1.9E+06 | 0.36 μs |
| *sps* →*spa* $k_7$ | 1.0E-08 | 2.1 yrs |
| *sas* →*saa* $k_8$ | 9.3E-09 | 2.4 yrs |

Our computed rate constants and derived half-lives at 200~K are shown in Table 1, which shows that $NH_2$ group flips are extremely fast while O–H dihedral rotamerisations are some $10^{10}$ times slower, while H-atom transfer is slower yet again by a similar factor with quantum-mechanical tunnelling playing the dominant role in both cases. The gas-phase value for the half-life for *aaa* → *saa* of 6.4 ms is much faster than the cryogenic experiments of ~4 s showing that reactions on noble-gas matrices are generally slowed by molecule–matrix interactions (Mandelli, 2026).

Recall that only two conformers have been studied by microwave spectroscopy in the laboratory. These experiments were performed either by heating solid glycine to 150–180ºC in a static cell (Ilyushin, 2005) or by vaporising glycine in a heated reservoir nozzle at 170ºC to generate a supersonic pulsed beam (Lovas, 1995). Under these conditions, spectra of *saa* and *ass* were obtained; although the thermal energy of ~3.7 kJ $mol^{-1}$ at ~440K is sufficient to populate more than one high-energy conformer, only *ass* can survive long enough to be spectroscopically characterised.

At 450 K only three higher-energy conformers are expected to be populated: *ass* (+3.7 kJ $mol^{-1}$) *spa* (+5.8 kJ $mol^{-1}$) and *sas* (+6.8 kJ $mol^{-1}$). The *spa* conformer decays immediately, while *sas* converts to *saa* via H-atom transfer. Although this process is much slower, with a half-life of 4.6 s, it would nevertheless be difficult to sustain a significant concentration of the *sas* conformer in the beam.

# Conclusions

We conclude that if all the conformers are initially present in the interstellar medium, only the two lowest-energy forms, *ass* and *saa*, survive long enough to be detectable. Because the dipole moment of *ass* is about five times larger than that of *saa*, it has a clear observational advantage. Laboratory spectroscopic efforts should therefore focus on refining and extending its spectral coverage, in conjunction with that for *saa*, to maximise the chances of detecting glycine in the ISM.

## Acknowledgements

We thank the Irish Centre for High-End Computing (ICHEC) for the provision of resources.

## Author Contributions

Authors Dr. Judith Wurmel and Prof. John M Simmie contributed in equal measure to the article.

## Statements and Declarations

Not applicable.

## Funding

This research did not receive any funding.